\documentclass[a4paper,12pt]{article}
\usepackage{graphicx} 
\usepackage{epstopdf}

\usepackage{setspace}
\begin{document}


\title{\bf Looking for the Signals of the Missing Baryons in the Extragalactic Background Light}
\author{ Wei Zhu$^1$\footnote{Corresponding author, E-mail:wzhu@phy.ecnu.edu.cn},
and Rong Wang$^2$\\
\normalsize $^1$ Department of Physics, East China Normal University, Shanghai 200241, China \\
\normalsize $^2$ Institute of Modern Physics, Chinese Academy of Sciences, Lanzhou 730000, China \\
}

\date{}

\newpage

\maketitle

\vskip 3truecm

\begin{abstract}

The missing baryons in the universe are assumed to be hidden in the whole space as a warm-hot
intergalactic medium (WHIM). Finding them is one of the important subjects in modern cosmology.
In this paper, we point out that the very high energy electron/positron rays may light up the WHIM
due to the anomalous bremsstrahlung according to an improved Bethe-Heitler formula.
The resulting excess of the extragalactic background light (EBL) can be observed by the direct measurement method.
A possible explanation on the difference between the direct and indirect measurements of EBL is also proposed.
Thus, we find a new window to probe the WHIM properties via the EBL.

\end{abstract}

{\bf keywords}:  Missing Baryons; Improved Bethe-Heitler Formula; Extragalactic Background Light

\section{Introduction}

    The standard cosmological model (SCM) considers that most $(95\%)$ of the universe is composed of
dark energy $(70\%)$ and dark matter $(25\%)$, while the universe's remaining $5\%$ is the visible matter
partly in the form of stars, planets, interstellar dust...
However these observed matter so far can only account for half of
the visible matter in the SCM predictions (see Refs. 1 and 2 for details).
It implies that at least $1/2$ of the baryons are now ``missing''.$^{3,4}$
Cosmological simulations show that these ``missing''
baryons in the intergalactic medium may condense into a filamentary web
and undergo shocks that heat the baryons up to the temperatures $T\sim 10^5-10^7K$.$^{5,6}$
Finding this warm-hot intergalactic medium (WHIM), which contains the missing baryons, is crucial to validate the SCM.
The WHIM is hot and highly ionized, so that it could absorb or emit far-ultraviolet and soft X-ray photons.
However, these characteristic lines are formed in the atomic bound state.
This spectral method is invalid if the WHIM is too hot to show up in the Lyman-absorption,
or too diffused to emit the detectable X-rays.$^7$
Recently, the dispersion of $X$-ray afterglows from fast radio bursts (FRBs)
is used to prove the existence of the WHIM.$^{8,9}$
Although the preliminary results from these studies tend to support the WHIM model,
the conclusion has a large uncertainty since the FRB sources were produced occasionally
and their positions in the universe are undetermined.
Still, all mentioned measurements above are not enough
to give a complete physical picture of the WHIM.

    The purpose of this article is to try to find the possible signals
of the missing baryons in the visible light band,
which relates to the extragalactic background light (EBL).
The EBL is the diffuse background radiation accumulated
over the cosmic history from ultraviolet (UV) to infrared (IR) wavelengths.
A peak in the EBL spectrum around $1~\mu m$ is formed
by the direct emission from stars. Measurements of both the EBL spectral intensity
and its evolution are important to study both the star formation and the galaxy evolution processes.$^{10}$
We find that the cosmic rays may light up the WHIM and lead to the observed excess of the EBL.

    Direct measurements of the optical EBL are made by observing the total sky
brightness and with the measurements or modeling of all the foreground components.$^{11,12}$
In this method, these foreground components must be very accurately known.
However, the bright zodiacal light (ZL), which is created by the scattered sunlight off
the interplanetary dust particles, may arise a large uncertainty for the weak EBL intensity.

    To get rid of this problem, the indirect techniques were developed:
the counting galaxies and the gamma-ray attenuation measurements.$^{13}$
These indirect measurements are complementary for the direct measurements
and they are important to constrain the large uncertainty of EBL.
At the same time, the direct surface photometry measurements have been improved.
For example, the detectors are set at the vantage points, where the Earth's atmosphere
and the light from interplanetary dust are not the appreciable components of the diffuse sky brightness.
These researches show an excess brightness above the indirectly measured EBL intensity.
In particular, the NASA's New Horizons spacecraft is presently
over 50 AU away from the sun, where the ZL almost disappears.
A recent report from New Horizons shows a new component of unknown origin of EBL of $8.06\pm1.92~nW m^{-2}sr^{-1}$.$^{14}$
This discrepancy is interesting because it might point to the presence
of a truly diffuse emission component in the universe that is not resolvable into point sources.

    Let us imagine a given small range in the sky,
where very high energy (VHE) cosmic electrons and positrons interact
with the Coulomb field in the WHIM and emit a lot of soft bremsstrahlung photons.
According to quantum electrodynamics (QED), the contribution of these soft photons to the EBL intensity
is only $n_{\gamma}\sim 10^{-20}~nWm^{-2}sr^{-1}$ (see Sec. 3), which is much smaller than the observed $n_{\gamma}\sim 10~nWm^{-2}sr^{-1}$.
Therefore, the contributions of the normal bremsstrahlung photons can be neglected.

    However, the above conclusion should be reconsidered if we use an improved Bethe-Heitler formula.$^{15}$
The reason is as follows.  The electromagnetic shower is a common phenomenon in astronomy,
which comprise the bremsstrahlung and the pair production
in the nuclear Coulomb field (Figs.. 1b and 1c). The quantum theory of the electromagnetic shower
is based on the Bethe-Heitler formula,$^{16}$ which has been widely applied in many branches in physics including astrophysics.
However, the recent work [15] pointed out that the Bethe-Heitler formula should be modified to
a new form in the dilute ionized gas, where the target recoil can be neglected for the incident VHE cosmic rays.
In this case both the bremsstrahlung and the pair production
cross sections have an unexpected big increment, as
$\sigma_{brem}\sim 1/m^2_e$ in the Bethe-Heitler formula Eq. (2.1) is replaced by $\sigma_{brem}^{MD}\sim 1/\mu^2$
in the modified version Eq. (2.2) and the screening length (or the Debye length) is $\mu^{-1}\gg m_e^{-1}$.
This leads to a big screening-length effect, which was neglected for a long time.

    The ionized gas of extreme low density is an ideal place
for testing the anomalous bremsstrahlung and pair production
since the ionized atoms have a large screening length.
The nuclear Coulomb potential may spread into such a broad space,
where the bremsstrahlung or pair production induce almost no recoil of the target
since the incident high energy electrons or photons are far away from the nucleus.
This result can explain the strange difference among the precise measurements of cosmic electron-positron fluxes
at the GeV-TeV band by Alpha Magnetic Spectrometer (AMS), Fermi Large Area Telescope (Fermi-LAT),
DArk Matter Particle Explorer (DAMPE) and Calorimetric Electron Telescope (CALET).$^{17}$
We find that the anomalous cross section has about eight-orders
of magnitude higher than the normal cross section in this example.

    We will discuss the corrections from the anomalous bremsstrahlung to
the directly measured EBL intensity in Sec. 3.
We find that the distribution of the EBL intensity will be moved upward if adding the bremsstrahlung contributions.
The computed result is close to the observation data.

    Then we discuss the anomalous pair production in the indirect measurement method using gamma-ray
attenuation with the same theoretical framework in Sec. 4.
We noticed that the data of gamma-ray attenuation coincide with that of the counting galaxies.
However, the processes in Figs. 1b and 1c do not contribute to the counting galaxies
since the WHIM is diffusely distributed in the space.
Therefore, we consider that the screening length $\mu^{-1}$ takes a smaller value in the gamma-ray attenuation method.
We will give an explanation for that in Sec. 5.
We find a strong dependence of the directly measured EBL intensity on the screening length
in the electronic structure of the WHIM.
This result may open a new window to expose the electronic properties of the WHIM during the universe evolution.

\section{Improved Bethe-Heitler formula}

    We sketch and review some results of the improved Bethe-Heitler formula from the work [15].
When charged particles scatter off the electric field of proton or nucleus,
they can emit real photons in such interactions.
This is the famous bremsstrahlung (braking radiation).
A quantum-mechanical description of bremsstrahlung emission at the Born approximation is the Bethe-Heitler formula.
This formula for the high-energy incident electrons and at the soft photon limit reads
$$\sigma_{brem}(E,\omega)d\omega=\frac{4\alpha_s^3}{m^2_e}\ln\frac{2E}{\mu}\frac{d\omega}{\omega}, \eqno(2.1)$$
where $\mu^{-1}$ is the screening length. Equation (2.1) presents a strong reduced bremsstrahlung cross section
since $1/m^2_e$ is much smaller than the geometric cross section $1/\mu^2$ of atom, where $\mu^{-1}\geq 10^{-8}~cm$.

    The work [15] proved that if the target recoil can be neglected for a VHE incident electron,
the Bethe-Heitler formula should be modified as,
$$\sigma^{MD}_{brem}(E,\omega)d\omega=\frac{4\alpha_s^3}{\mu^2}\ln\frac{2E}{\mu}\frac{d\omega}{\omega}. \eqno(2.2)$$
It predicts a big enhanced bremsstrahlung cross section.
This is called the anomalous bremsstrahlung effect.

    The deeply ionized gas in the cosmic space is an ideal place
to see the anomalous bremsstrahlung effect since the dilute ionized atoms
have a large screening length, where the nuclear Coulomb potential may spread
into such a broader space, and the recoil of the bremsstrahlung event
can be neglected since the incident high energy electrons are far away from the nuclei.
Bremsstrahlung under this environment exhibits a big cross section.
On the other hand, the bremsstrahlung probability also decreases with the reduced density.
The above two opposite effects lead to a limited kinematic window,
where the bremsstrahlung cross section becomes anomalous.
According to the analysis of the work [17], the TeV-band of electron energy is such a possible window.

    The similar argument also satisfies for the electron-positron pair creation from the real high energy photon.
We consider a high energy photon traversing the atomic Coulomb field.
This photon has a probability of transforming itself into a pair of electron-positron.
The corresponding cross section in the leading approximation is given by,
$$d\sigma_{pair}=\frac{\alpha^3}{m_e^2}\ln\frac{4\omega^2}{\mu^2}(1-z)[(1-z)^2+z^2]dz, \eqno(2.3)$$
where $z=E_e/\omega$. In a situation where the target recoil can be neglected,
we have the anomalous pair production cross section,
$$d\sigma_{pair}^{MD}=\frac{\alpha^3}{\mu^2}\ln\frac{4\omega^2}{\mu^2}(1-z)[(1-z)^2+z^2]dz \eqno(2.4)$$

    The above improved Bethe-Heitler formula has been used to explain the difference of VHE electron-positron fluxes
in the ionosphere about $400\sim 500~km$ height above the Earth ground,
where the oxygen atoms are strongly ionized
and have extremely dilute density.

    We take AMS-Calet-DAMPE-FermiLAT data to test the improved  Bethe-Heitler formula since their measurements
are on orbit. There are other data of cosmic electron-positron flux, for example, HESS  and ATIC are ground-based measurements and balloon observations. They relate to the propagating models of electrons in atmosphere and both them have a large uncertainty. Therefore, we did not use them.

    The anomalous bremsstrahlung effect relates to the degree of ionization and density of atoms. We only give an
order of magnitude estimation of the screening length, which is much larger than the atomic scale. This is consistent with the characteristics of the physical environment in plasma. On the other hand, at altitudes ~400-500 km the plasma density and ionization degree are dependent from solar activity considerably. It would be result to lepton flux variations with time. These need to be further tested.

\section{Anomalous bremsstrahlung in the WHIM}

    The direct measurement of the EBL intensity was carried out by the special camera in the space.
After the subtraction of the contributions from the ZL and other foreground components,
the data show an obvious excess of EBL photons compared to the indirectly measured EBL intensity.
This indicates the presence of an unknown diffuse emission component.
However, the conclusion remains uncertain
since the estimations of the ZL contributions have large uncertainties.

    The NASA's New Horizons spacecraft is an excellent platform
for the EBL observation since it is presently over 50 AU away from the sun,
where the zodiacal light is almost disappeared.
A recent report from the New Horizons confirms an EBL component of unknown origin
of $n_{\gamma}=8.06\pm1.92~nW m^{-2}sr^{-1}$.$^{14}$
Although some of the authors have argued that the EBL includes a substantial component of light
from stars tidally removed from galaxies, or a population of faint sources in extended halos.
None of these hypotheses may be correct.$^{14}$
We try to explain this excess of the EBL photons and to point out that
it could be arisen from the contributions of the anomalous bremsstrahlung
when the VHE cosmic electron/positron fluxes pass through the WHIM.

    We consider a small range in the deep space, where the incident VHE cosmic electron-positron rays
interact with the Coulomb field of the WHIM and emit soft bremsstrahlung photons.
The screening length in Eq. (2.2) depends on the ionization state
and spatial distribution of the missing baryons in the WHIM.
If the probing range is small enough, the resolution of camera can not distinguish the WHIM structure.
Thus, we take an average screening length $\mu^{-1}$ in Eq. (2.2) to describe the probed WHIM.
The contributions of the anomalous bremsstrahlung to the EBL intensity is
$$n_{\gamma}(\omega)d\omega =\int_{E_{min}}^{E_{max}} n_{\gamma}(E,\omega)dEd\omega
=\int_{E_{min}}^{E_{max}} J_e(E)P(E,\omega)dEd\omega. \eqno(3.1)$$
where $J_e(E)$ is the electron-positron flux at energy $E$.
We take $E_{min}=0.1~TeV$ and $E_{max}=10~TeV$ according to the Ref. [17].
$P(E,\omega)d\omega/\omega$ is the photon number emitted by the electrons
in the range $\omega\rightarrow\omega+d\omega$.
The relation between the emission power spectrum $P(E,\omega)d\omega$ and the bremsstrahlung cross section is written as [18],
$$\sigma_{brem}^{MD}(E,\omega)d\omega=\frac{P(E,\omega)d\omega}{n_pv_e\omega}, \eqno(3.2)$$
where $n_p\sim 1/m^3$ is the average density of the missing baryons in the WHIM and $v_e\simeq c$.

    We use the parameterized distribution function
$$J_e(E)=10^{-2}\left(\frac{E}{100GeV}\right)^{-2}[m^{-2}s^{-1}sr^{-1}], \eqno(3.3)$$
to model the electron-positron flux in the energy range $100~GeV<E<1~TeV$$^{17}$
and assume that this estimation is available in the intergalactic space, although
there is a large uncertainty of the spectra index in the TeV range.

    Note that the azimuth angles in both $n_{\gamma}(\omega)$ and $J_e(E)$ are almost coincide,
since the soft photons are concentrated in the moving direction of the high energy electrons.
Thus, we have,
$$n_{\gamma}(\omega)=\int_{E_{min}}^{E_{max}}J_e(E)\frac{4\alpha_s^3}{\mu^2}\ln\frac{2E}{\mu}n_pv_edE.\eqno(3.4)$$
This result shows a flat distribution and its height sensitively depends on the value of $\mu$.
The intergalactic electron intensity may be different from Eq. (3.3), but it will not cover up a large anomalous bremsstrahlung
effect.

    In Fig. 2 we show our resulting $n_{\gamma}(\omega)\sim \lambda$ (dashed curve)
with $\mu=10^{-15}~GeV$ or $\mu^{-1}=10~cm$.
Dotted curve is a global fitting of the EBL intensity data from the indirect method.$^{19}$
The sum of the above two components (solid curve) is consistent with
the data from New Horizons spacecraft (solid point).$^{14}$
We also present some other directly measured EBL intensities (all other points),$^{20-22}$
where the ZL contributions are deducted by using the special models rather than the measurements in the observations.
Although our estimation is still preliminary, one can find that
the contribution from the anomalous bremsstrahlung to the EBL density seems possible.
On the other hand, the large excess around $\lambda\sim 1~\mu m$ was suspected to be from
the highly redshifted ultraviolet radiation of the first generation of stars.
However, the evolution models based on the observations of high-redshift galaxies indicate that
the brightness caused by first stars are much lower than the observed excesses.$^{23}$
The increment of the EBL due to the anomalous bremsstrahlung may compensate this deficiency.

    Note that the solution of Eq. (3.4) can not be applied without restrictions.
The applications of the anomalous bremsstrahlung requests very low photon energy
so that neglecting the recoil effect is safe.
On the other hand, the radiation corrections are needed when the photon energy $\rightarrow 0$ in Eq. (3.4).$^{24}$
The infrared cutoff should be used to stop the infinite increasing of $\sigma^{MD}\propto 1/\omega$
due to the negative corrections of the virtual processes.
The radiation corrections depend on the experimental conditions (energy and angular resolutions).
Thus, we assume that the intensity $n_{\gamma}\propto \omega \sigma^{MD}$ is suppressed when the photon energy $\omega$
is below the resolution of the detector. Therefore, we restrict the solution of Eq. (3.4) to be
in the range of $\lambda \approx 0.1-3~\mu m$, which is shown in Fig. 2.

    According to the Bethe-Heitler formula (2.1), Eq. (3.4) should be replaced by
$$n_{\gamma}(\omega)=\int_{E_{min}}^{E_{max}}J_e(E)\frac{4\alpha_s^3}{m_e^2}\ln\frac{2E}{\mu}n_pv_edE,\eqno(3.5)$$
and we have $n_{\gamma}\sim 10^{-20}~nWm^{-2}sr^{-1}\ll 10~nWm^{-2}sr^{-1}$.
This is why the contributions of the normal bremsstrahlung photons are neglected.

\section{Anomalous pair production in the WHIM }

    An indirect measurement of the EBL intensity is based on the absorption of gamma rays
via the pair production off real EBL photons. This effect leads to the distance-dependent
suppression of the gamma-ray flux from active galactic nuclei (AGNs).$^{10,13}$
AGNs are the extragalactic objects characterized with extremely luminous electromagnetic radiations.
The missing baryons in the WHIM may deform the gamma-ray distributions
due to the anomalous pair production process described by Eq. (2.4).
We begin from a case without the WHIM.
The EBL may modify the propagation of VHE gamma-rays traveling
through intergalactic space due to pair production in the channel
$\gamma_{VHE}+\gamma_{EBL}\rightarrow e^-+e^+$ (Fig. 1a).
The observed VHE gamma-ray flux on the Earth is related to the initial flux at the source as,
$$\Phi_{\gamma}^{ob}(E_{\gamma})=\Phi_{\gamma}^{in}(E_{\gamma})e^{-\tau_{\gamma\gamma}},\eqno(4.1)$$
where the optical depth $\tau_{\gamma\gamma}$ relates to the pair production cross section
of two photons $\sigma_{\gamma\gamma}$ and the density $n_{\gamma}$
of the soft photon in the EBL as,
$$\tau_{\gamma\gamma}(E_{\gamma})
=\int_0^R\int_{\lambda_{min}}^{\lambda_{max}}\sigma_{\gamma\gamma}(E_{\gamma},\lambda)n_{\gamma}(
\lambda, r)d\lambda dr,$$
$$~~~~around~E_{\gamma}\sim 1~TeV. \eqno(4.2)$$
This is the foundation of the indirect method of the EBL measurement.

    If we considering the WHIM, the contributions of pair production in the nuclear Coulomb field
of the WHIM (figure 1c) will be added to the observed optical-depth data in Eq. (4.2).
Therefore, the measured optical depth based on the pair production model should include
the contribution $\gamma+p\rightarrow e^++e^-$ due to the WHIM,
i.e., $\tau_{\gamma\gamma}\rightarrow \tau_{\gamma\gamma}+\tau_{\gamma p}$.
The optical depth in the WHIM is evaluated with the anomalous pair production as,
$$\tau_{\gamma p}(E_{\gamma})=\int_0^R\sigma_{pair}^{MD}(E_{\gamma},r)n_p(E_{\gamma},r) dr,\eqno(4.3)$$
where $n_p$ is the proton density in the WHIM.

    We have$^{24}$ $\sigma_{\gamma p}\sim 0.1\sigma_{\gamma\gamma}$ around $E\sim 1~TeV$
with $n_p/n_{\gamma}\sim 10^{-9}-10^{-10}$ and $\sigma_{pair}^{MD}\sim (m_e/m_{\mu})^2\sigma_{Pair}$.
To have the similar magnitude of gamma-ray attenuation from the WHIM
as that from the EBL, $\tau_{\gamma p}\sim \tau_{\gamma\gamma}$,
the screening length of the WHIM is required to be,
$$\left(\frac{m_e}{\mu}\right)^2=10^{10}-10^{11}. \eqno(4.4)$$

    Note that the counting-galaxy method is irrelevant to the WHIM structure
since the virtual photons in the Coulomb field are not counted in this method.
The EBL data deduced from the counting galaxies and the gamma-ray attenuation method
coincide with each other.$^{13}$
Therefore, the correction to the optical depth
from the anomalous pair production on the WHIM should be neglected.
This implies that $(m_e/\mu)^2\ll 10^{10}-10^{11}$,
or $\mu^{-1}\ll 10^{-5}~cm$ in Eq. (4.3).
We will discuss these results in the next section.

\section{Discussions}

    Both the EBL and the WHIM are the products of the universe evolution,
and they convey some important information of the cosmic formation history.
Many works studied the evolution of the optical depth $\tau_{\gamma\gamma}$.$^{12,13}$
In this work, we focus on the evolution of the properties
of the WHIM with the redshift, based on the optical-depth correction
from the WHIM with the improved Bethe-Heitler formula.

    The screening length $\mu^{-1}$ is sensitively dependent
on the ionization status of the missing baryons in the WHIM.
We take two extreme examples to illustrate it.
(i) If the missing baryons exist in the form of neutral atoms,
we have $\mu^{-1}\sim 10^{-8}~cm$ or $n_{\gamma}\sim 10^{-15}~nW m^{-2}sr^{-1}$
calculated with Eq. (3.4);
(ii) The missing baryons and electrons are uniformly distributed in the space as the plasma form,
where the maximum of the average screening length is
$\mu^{-1}=(1/n_p)^{1/3}\sim  10^2~cm$,
and we have $n_{\gamma}\sim 10^5~ nWm^{-2}sr^{-1}$.
These two examples show that as we use the average screening length
to describe the ionization property of the atom gas in the WHIM,
its value has a broadly reasonable range from the beginning of re-ionization
(i.e., $\mu^{-1}>10^{-8}~cm$) to a maximally ionized state of the baryon-electron plasma.

   The bremsstrahlung energy loss relates to the elemental QED processes and a concrete ionization model. The later
may provides a process-dependent parameter $\mu$. In the above discussions we give a possible range of the parameter $\mu$ in the WHIM according to the EBL data. We noticed that there are numerous ionization models are devoted to describe bremsstrahlung energy loss
in interstellar medium.$^{25}$  It is necessary to improve our estimation using these models but based on the improved Bethe-Heitler formula. We will study them in future work. On the other hand, as work [15] have pointed out that the anomalous bremsstrahlung effect origins from a fact, that the screening parameter in a logarithmic factor of the cross section moves to a denominator position if the recoil effect can be neglected at the high energy approximation. We believe that the application of the ionization model will not inhibit a big correction of the anomalous bremsstrahlung in the WHIM, although the estimation may become more accurate.

    The standard cosmological model predicts that the missing baryons
in the universe evolution are concentrated in the WHIM
due to continuously shock-heated process.$^{1,2}$
Cosmological simulations find that the WHIM contains a large fraction
of the baryons at redshift $z\simeq 0$ in the form of highly-ionized gas.$^{26}$
Theoretically, an atom which is bounded in an infinite heavy web
can completely absorb the recoil effect.
However, this bounded atom can not avoid the obvious recoil corrections
due to the intense collisions at a short impact parameter.
Therefore, these bounded atoms in the WHIM do not have the anomalous bremsstrahlung
or the anomalous pair production.
We consider the atomic gas inside the WHIM at high redshift.
The direct EBL intensity is measured by using a long time exposition
of the camera at low redshift $z\sim 0$.
We have extracted a large screening length $\mu^{-1}\simeq 10~cm$ at $z\simeq0$ in Sec. 3.
It implies that a lot of atoms in the WHIM at $z\rightarrow 0$ are almost re-ionized
in the complete baryon-electron plasma state.
This result is consistent with the observations, which gives the evidence
that a significant portion of the baryons are located at $z\simeq 0$ in the WHIM.$^{27}$
On the other hand, the gamma rays in the indirect measurement of the EBL intensity
come from the AGNs at large redshift.
The contributions to the optical depth from highly-ionized gas in the WHIM at low redshift
only account for a small part of the total optical depth in Eq. (4.3).
Therefore, we have the average screening length $\mu^{-1}\ll 10^{-5}~cm$
for the anomalous pair production discussed in section 4,
which leads to the result $\tau_{\gamma p}\ll \tau_{\gamma \gamma}$.
Thus, we explained why we use two different values of the parameter $\mu^{-1}$ in Sec. 3 and Sec. 4.

    Usually, the absorptions or emissions of far-ultraviolet and soft X-ray photons.$^{28}$
or the hyperfine transition of neutral hydrogen (called 21-cm line)$^{29}$ are
used to look for the signals of the missing baryons in the WHIM.
However, the WHIM is extremely difficult to be observed, for its tremendously
low density and high temperature are believed to elude the detections of the absorption and emission signals.
Therefore, the properties of the WHIM at $z=0$ is still elusive with the traditional detecting methods.
Comparing with the characteristic spectra of the WHIM,
the sensitive dependence of the improved Bethe-Heitler cross section
on the average screening length provides a new method to complement
the study of the WHIM structure.

    In summary, we try to establish a connection between the WHIM and the EBL.
The found relation answers the following cosmological questions:
Where are the missing baryons?
Is the sky absolutely dark if all star lights are removed?
According to an improved Bethe-Heitler formula, we find that
the WHIM may be lighted up by the VHE cosmic electron/positron fluxes,
and this effect can be observed by the directly measured EBL intensity.
Using a sensitive relation between the EBL intensity and the WHIM electronic structure,
the improved Bethe-Heitler formula may open a new window to study the WHIM structure.

\section{ Acknowledgments}
This work is partly supported by the National Natural Science Foundation of China No.11851303 and No.12005266.

\newpage

\begin{figure}
  \begin{center}
   \includegraphics[width=0.8\textwidth]{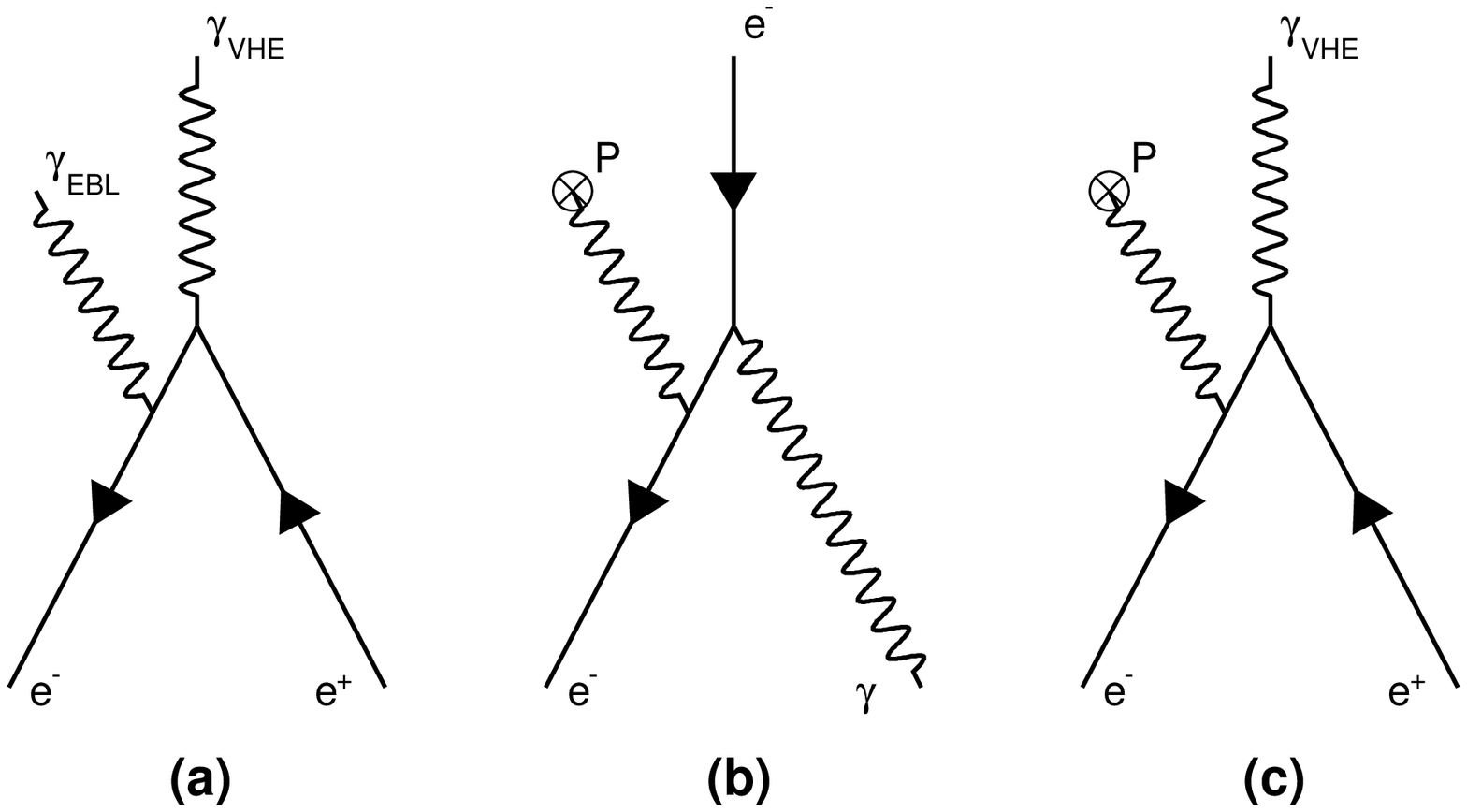} 
   \caption{(a) The interaction of high energy photon with soft photon in the EBL.
            (b) and (c) Bremsstrahlung and pair creation in the nuclear Coulomb field,
            which can be described with an improved Bethe-Heitler formula
            in the large-scale medium of dilute ionized matter.
}\label{fig:1}
  \end{center}
\end{figure}

\begin{figure}
  \begin{center}
   \includegraphics[width=0.8\textwidth]{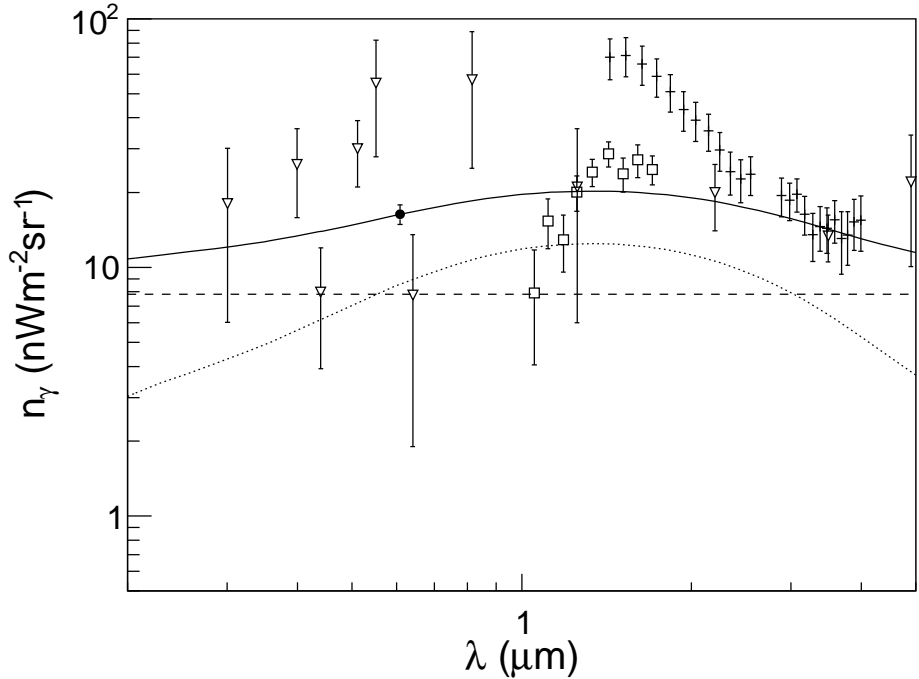} 
   \caption{The contribution from the anomalous bremsstrahlung (Fig. 1b)
   to the EBL intensity is presented with the dashed line.
   Dotted curve is taken from,$^{19}$ which is fitted to the indirectly measured EBL intensity.
   Solid curve is the sum of the above two components. The filled circle with
   error bar is a recent data from New Horizons,$^{14}$ which does not have the ZL component.
   All other data points are taken from some different direct EBL measurements$^{20-22}$
   but with the model-dependent corrections for ZL.
}\label{fig:2}
  \end{center}
\end{figure}

\end{document}